# Estudo de Ciclos Multidecadais nos Índices Climáticos Usando a Análise de Wavelet para o Norte/Nordeste do Brasil
## Multidecadal Cycles Study in the Climate Indexes Series Using Wavelet Analysis in North/Northeast Brazil

Cleber Souza Corrêa[1]; Roberto Lage Guedes[1]; Karlmer Abel Bueno Corrêa[2] & Felipe Gustavo Pilau[2]

[1]Instituto de Aeronáutica e Espaço (IAE),
Praça Marechal do Ar Eduardo Gomes, 50, Vila das Acácias, 12228-904 São José dos Campos, São Paulo, Brasil
2 Universidade de São Paulo, Escola Superior de Agricultura "Luiz de Queiroz", AGRIMET, Departamento de Engenharia de Sistemas Agrícolas - ESALQ/USP, Av. Pádua Dias, 11, 13418-900 Piracicaba, São Paulo, Brasil
E-mails:clebercsc@fab.mil.br; roblg331@gmail.com; karlmerabc@usp.br; fgpilau@usp.br

**Resumo**

Este estudo investiga as séries temporais dos índices climáticos, com um período dos 80 anos, os mais recentes, com valores mensais, utilizando o índice da Oscilação Decadal do Pacifico (ODP), o Índice de Oscilação Sul (SOI) e a série temporal mensal da atividade solar com o número de Manchas Solares (MS), obtidas no site do Serviço Nacional de Meteorologia - Centro de Previsão Climática e dados de manchas solares do Data Center Mundial SILSO, Observatório Real da Bélgica, Bruxelas. O software estatístico R foi utilizado com o pacote WaveletComp para gerar os espectros de potência de waveletMorlet, análise bivariada com wavelet cruzada e o uso do pacote biwavelet. Como resultado, os ciclos predominantes foram obtidos com variabilidades de ordem de 32, 64, 128 e 256 meses, aproximadamente 2,66, 5,33, 10,66 e 21,33 anos. Essas frequências são observadas na variabilidade do período de janeiro de 1933 a setembro de 2016, no total de 993 meses (82,75 anos), caracterizando variações decadais para multidecadal. Esses ciclos multidecadais observados (da ordem de 10,66 e 21,33 anos podem mostrar um certo grau de associação com a variabilidade da atividade solar (Mancha solar) com os ciclos de variabilidade climática no sistema Atmosférico/Oceânico. Dados da precipitação total de janeiro de 1951 a setembro de 2017 foram analisados, cujas informações foram obtidas nos aeroportos de Belém, São Luiz, Fortaleza, Natal e Fernando de Noronha, caracterizando uma diagonal do norte para o nordeste brasileiro, observando que essas séries apresentaram similaridade nos ciclos decadais e multidecadais encontrados no SOI/PDO e MS.
**Palavras-chave:** Análise de séries temporais; Decadal para ciclos Multidecadais; Análise Wavelet

**Abstract**

This study investigates the climatic indexes time series, with a period of the most recent 80 years, with monthly mean values, using the Pacific Decadal Oscillation Index (PDO), Southern Oscillation Index (SOI), and the monthly solar activity time series with the sunspots number (MS), obtained from the National Weather Service's website - Climate Prediction Center and Sunspot data from the World Data Center SILSO, Royal Observatory of Belgium, Brussels. The statistical software R was used with the WaveletComp package to generate the Morlet wavelet power spectra, bivariate analysis with cross-wavelet and the biwavelet package. As a result, predominant cycles were obtained and order variabilities of 32, 64, 128 and 256 months were observed, with approximately 2.66, 5.33, 10.66 and 21.33 years. These frequencies are observed in the variability of the period from January 1933 to September 2016, in the total of 993 months (82.75 years), characterizing decadal variations for multidecadal. These observed multidecadal cycles (of the order of 10.66 and 21.33 years may show a degree of association with the variability of solar activity (Sunspot) with the climatic variability cycles in the Atmospheric/Ocean system. Data from total rainfall from January 1951 to September 2017 were analyzed information located at Belem, São Luiz, Fortaleza, Natal and Fernando de Noronha airports, characterizing a diagonal from the north to the Brazilian northeast, observing that these series show similarity in the decadal and multidecadal cycles found in the SOI/PDO and Sunspot series.
**Keywords:** Time-series analysis; Decadal to multidecadal cycles; Wavelet analysis

## 1 Introduction

The temporal scales involved in ocean/atmosphere interaction studies representing the Southern Oscillation Index variability and the El Niño-Southern Oscillation (IOS-ENSO) have estimates of time scale from 2 to 5 years, such behavior reinforces a well-defined biannual oscillation characteristic. There are studies that evaluate the Southern Oscillation Index variability, which has a biannual signal variation component, (Ropelewski & Jones, 1987; Elliott & Angell, 1988; Rao & Hada, 1990; Kawamura *et al.,* 1995; Grimm *et al.,* 1998; Kayano & Andreoli, 2006; Capotondi *et al.,* 2015; Astudillo *et al.,* 2016). Solar activity has a direct influence on many terrestrial systems, the relationship between the solar energy net balance that reaches the earth's surface and especially the oceans directly affect the climate cycles in the atmosphere and in the terrestrial/oceanic system.

There are some scientific studies that relate this solar activity influence to the annual climatological cycle and to physical and temporal scales with interannual and decadal variability, suggesting that the solar-related signals affect the temperature on the ocean, and with its temperature being higher and being able to reach and penetrate to 80-160m deep in the oceans, resulting in direct influence on the climatic series behavior and their scales with decadal variability (Ellias *et al.,* 1978; Levitus *et al.,* 2000; Willis *et al.,* 2004 & Levitus *et al.,* 2012).

Lassen & Friis-Christensen (1995), showed that the duration of the solar cycle in the last five centuries was associated with the Earth's climate, with a well-defined activity with an eleven-year cycle in the sunspots number. Since solar activity in approximately 11 years ranged from 8 to 17 years within an 80-90 years period.

*In* Rasmuson *et al.* (1990), using surface data for sea wind and sea surface velocity from the period 1950-1987, together with sea-surface temperature and sea-level pressure data from several stations in the Pacific, were two dominant time scales of El Niño-Southern Oscillation (ENSO) variability were identified. One is a biennial mode, with periods close to 24 months, and the other with a concentration variance with lower frequency in periods from four to five years. A well-defined biannual component of the ENSO variability, strictly closed with the annual cycle, appears in the surface wind field of the eastern equatorial Indian Ocean. This is part of a biennial circulation on a larger scale relative to the eastern low latitude sector of the Indian Ocean-Western Pacific. This biennial circulation is an ENSO variability key element. It exhibits a strong tendency to block phase with the annual cycle and introduces a certain regularity degree in the ENSO cycle.

Astudillo *et al.* (2016), show that the information contained in IOS is enough to provide nonlinear attractor information, allowing the detection of predictability for more than one year: 2, 3 and 4 years in advance throughout the record with an acceptable error. This is possible because the low frequency variability of the IOS presents positive long-term autocorrelation.

There are studies that refer to the IOS/ENSO variability with the Pacific Decadal Oscillation variability, (Newman *et al.,* 2003; Rodgers *et al.,* 2004; Zhang & Church, 2012; Gu & Adler, 2013; Kucharski *et al.,* 2015), aiming at understanding the dynamic structure that controls decadal variability in the Pacific Ocean and its interactions with climate change on a global scale.

Stauning (2011), shows that the solar activity ratio to climate on earth is magnified with the latest temperature and sunspots data series. There is a strong correlation between solar activity and terrestrial temperatures delayed in 3 years. A regression analysis between the solar activity represented by the mean number of sunspots in the cycle and global temperature anomalies, averaged over the same interval but delayed by 3 years. It is suggested that the variations in the cycle and also the long-term variations in global temperatures during the 135 years examined are mainly caused by corresponding changes in the total level of solar irradiation that represent the energy production of the solar core but are still modulated by energy variable transmission properties in the outer Sun regions.

Andreoli *et al.* (2004), observed in their work that there existed variations in precipitation in For-taleza and sea surface temperature in the Pacific and Atlantic during 1856-1991 was analyzed using wavelet transform and partial correlations. Indices was obtained for precipitation anomalies and SST anomalies (SSTA) in the Niño-3 regions and Tropical Atlantic (DTNS index). The precipitation and DTNS indices contain strong variability at 11-14 year scale. Maximum variances for the Niño-3 index occur at interannual scales, but modulated by 12-20 year oscillations. Therefore, the connections between SSTA in the Pacific east and precipitation in the NEB in the decadal scale can be a response to Niño-3 in the interannual scale.

Guedes *et al.* (2005), in their work analyzed the precipitation series in Fortaleza, observed that there are characteristics in the spectral signals, observed in the mentioned temporal scale, could in future works, be correlated with solar activity time series that, in turn, have 11 years cycles, with variations between 8 and 15 years.

This work seeks to analyze series with total period of the 80 years order, since the series are available, without failures nor some type of adjustment to estimate the average monthly values observed. The use of wavelet analysis seeks to capture the variability in space and time of the predominant frequencies in the time series in the eighty years, the most recent, allowing the intradecadal and multidecadal cycles observation, showing the synchronization between the time series signals analyzed and in these predominant cycles observed, showing the strong atmosphere/ocean and the solar activity interaction, it could influence the decadal and multidecadal cycles signal in the monthly total precipitation historical series over the Brazil north and northeast.

## 2 Methodology
### 2.1 Data Analyzed

Southern Oscillation Index (IOS) data were obtained from the National Prediction Center's Web site (www.cpc.ncep.noaa.gov/data/indices/). IOS data were used monthly and normalized from January 1933 to September 2016. The Pacific Decadal Oscillation Index (PDO) was derived as the main major component of the monthly anomalies of SST in the North Pacific Ocean at the 20N pole. The mean monthly SST anomalies are removed to separate this pattern of variability from any "global warming" signal that may be present in the data (Zhang *et al.*,1997; Mantua *et al.*,1997).

The PDO index data were obtained from the NOAA website (http://www.ncdc.noaa.gov/teleconnections/pdo/), the data.csv file with the index information was obtained, with approximately 162 years in the time series analyzed, with monthly values.

Significant monthly average data on the sunspots number (MS) were obtained at http://www.sidc.be/silso/datafiles (data from the Sunspot of the World Data Center SILSO, Royal Observatory of Belgium, Brussels), which was transferred the data file (SN_m_tot_V2.csv) with information, from 1749 to 2016 years with a 267 years series.

The PDO and Mean Sunspot data were adjusted so that the three files were from the same period from January 1933 to September 2016, in total there were 993 months with IOS value, PDO and average monthly Sunspots, about 83 years in the time series analyzed. The total monthly rainfall surface data were used from five meteorological stations of the Brazilian Air Space Control Department (DECEA) climate database. The data used were in the Belem, São Luiz, Fortaleza, Natal and Fernando de Noronha airports, the time series analyzed were from January 1951 to September 2017 with 67 years.

### 2.2 Wavelet Analysis

The WaveletComp is a R package that using for analysis based on univariate and bivariate time series wavelets. The WaveletComp version 1.0 (R Foundation for Statistical Computing), Roesch and Schmidbauer (2014) with the frequency analysis of uni and bivariate time series using Morlet, (Morlet *et al.* 1982a, 1982b; Goupillaud *et al.*,1984) and the biwavelet package by Grinsted *et al.* (2004). Morlet wavelet, in the version implemented in WaveletComp, uses equation (1):

$$\psi(t) = \pi^{\frac{-1}{4}} e^{iwt} e^{\frac{-t^2}{2}} \qquad (1)$$

The "angular frequency" ω (or rate of rotation in radians per unit time) is defined as the value 6, which is the preferred value in the literature, since the Morlet wavelet is solved approximately in an analytical way; and with an oscillation equals 2π (radians); therefore, the period (or the inverse frequency) measured in units of time is equal to 2π /6.

The Morlet wavelet transform of a time series ($x_t$) is defined as the convolution of the series with a set of "wavelet daughters" generated by the "mother wavelet" by time translation by τ and defining the scale by s, in equation (2):

$$Wave(\tau,s) = \sum_t x_t \frac{1}{\sqrt{s}} \psi^*\left(\frac{t-\tau}{s}\right) \quad (2)$$

with * denoting the complex conjugate. The position of the daughter wavelet in the time domain is determined by the location of the time parameter τ being displaced by dt a time increment. WaveletComp rectifies the wavelet power spectrum (cross-wavelet) according to Liu *et al.* (2007), in the univariate case, and Veleda *et al.* (2012), in the bivariate case, to avoid "biased" results in filtering and estimating the associated high frequencies or in the sense of variability estimates with short periods in the time series.

The analyzed phenomena and the time series would tend to be underestimated by conventional approaches. Implemented options for smoothing in period time and /or direction, it is necessary to perform the calculation with the Coherence wavelets methodology with their multiples, Liu (1994). The set choice of scales s determines the series wavelet coverage in the frequency domain.

## 3 Results

The results obtained by the WaveletComp package are presented in Figure 1 with graphs two types, Figures 1-A, 1-B and 1-C, are graphs with wavelet coherence analysis. Figure 1-A shows the IOS time series showing bands (order positive signal about 1.6) occurring between 32 months (± 2.66 years) and the lower 32 months with higher intensity, was observed by Stauning (2011) and Astudillo *et al.* (2016), obtained order results of 2 to 4 years of the signal analysis in the IOS time series, and with 64 months (± 5.33 years), already in the series middle for the final presents 128 months (± 10.66 years ) well-defined, already recognized in the literature and observed by Wang *et al.* (2015), observed a upper ocean heating decadal variability signal in the Pacific, responding to the solar cycle of 11 years. Figure 1-B represents the PDO time series representation, and a positive signal at the beginning of the series at 256 months (± 21.33 years) from the middle to the end with an intense signal, on the order of 0.7, of 128 months, and intense (≥ 0.7) with distributed 64 months to around 32 months.

Figure 1-C shows the temporal series of the monthly sunspot mean and presents an intense signal, of the order of 1,7, in 128 months and until about 200 months, and a signal of less intensity in 64 months throughout the series Sunspots and a slight signal in 256 months. The period of 20 to 27 years may be related to the work of Molion (2005), discusses the global mean temperature that the variability may be linked to Pacific Decadal Oscillation between a periodicity of 20 to 25 years. The signal of 256 months (± 21.33 years) appears in the PDO series and in the series of sunspots with lower intensity. Figures 1-D, 1-E and 1-F, as 2-A, 2-B and 2-C present the estimation information of the bi-variable Cross-wavelet calculation. Figures 2-A, 2-B and 2-C present the bi-variable cross-wavelet spectrum-time spectra, the spectra observed in the spectra are similar to the bands observed in the wavelet coherence figures, with periods of 2.66, 5, 33, 10.66 and 21.33 years, showing the similarity of solar activity cycles and the IOS and PDO indices.

These four periods define temporal ranges that are to be understood as ranges that in the extension of the temporal series can occur variations, as example the period of 2.66 years can represent variability of 2 to 4 years, as also the other periods represent ranges of variability in the last 80 years in the three-time series analyzed. Figure 2-b (PDO-IOS) shows a more pronounced signal in the cycle of 256 months, of the order of 21.33 years, also showing that the signal of 11 years (± 10, 66 years) decreases, which shows that internally the ocean/atmosphere interaction predominates with greater intensity in the signal of 64 months (± 5.33 years) between the PDO /IOS.

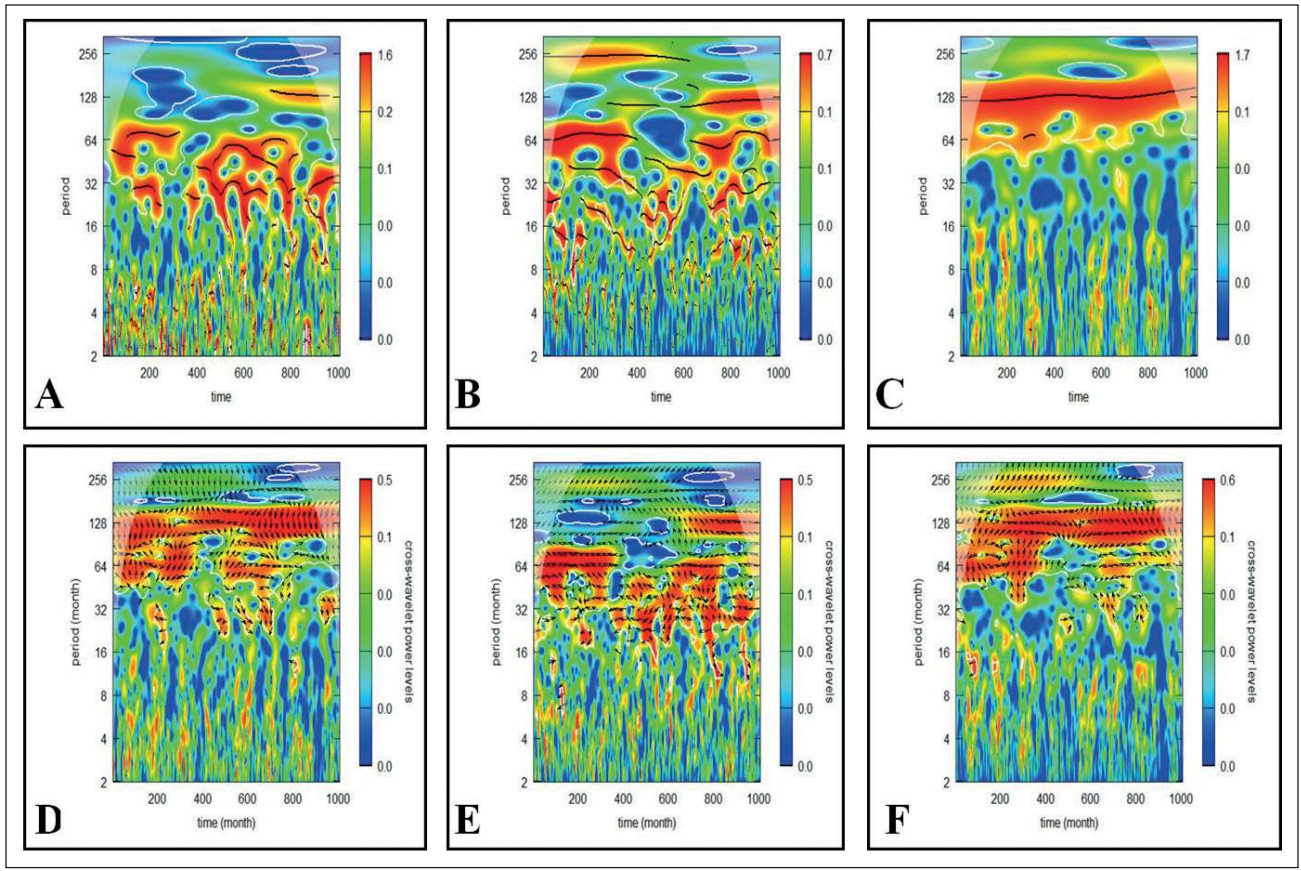

Figure 1 Time series relationship with bi-variable analysis using wavelet analysis, showing Coherence wavelet analysis, **A.** IOS; **B.** PDO and **C.** Sunspot. The bi-variable power spectrum with cross-wavelet, **D.** IOS-Sunspot; **E.** IOS-PDO and **F.** PDO-Sunspot.

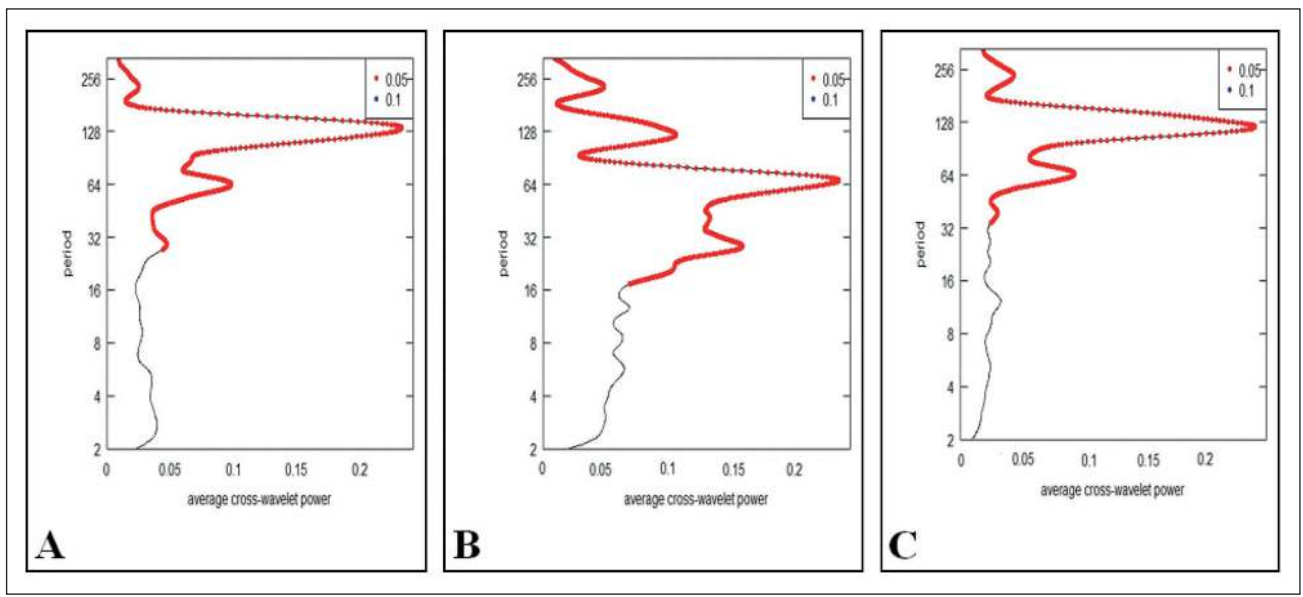

Figure 2 The cross-wavelet bi-variable mean-time spectrum, **A.** IOS-Sunspot; **B.** IOS-PDO and **C.** PDO-Sunspot.

The results found characterized a multidecadal variability with order time scales about 10.66 and 21.33 years. Figure 3-A and 3-E in the Belém and Fernando de Noronha localities show well-defined range cycles with 21.33 years. Well-defined 11 years periods appear in the monthly total rainfall series in the following locations: Belém, Fortaleza and Natal in the Figure 3-A, 3-C and 3-D. In Fernando de Noronha (Figure 3-E) presents a weak sign of 11 years but the time series shows well defined periods about 21.33 years. All the analyzed time series show a strong high frequency signal in the 12 months period. Figure 3-B shows the São Luiz locality, where it shows a weak sign of 64 months (± 5.33 years) as well as 11 years, but in all localities shows the sign of 5.33 years.

These results characterizing the low multidecadal frequency existence in the precipitation monthly total series for the north and northeast Brazil and the similarity with the solar activity cycles.

## 4 Conclusions

The statistical technique used was able to represent the predominant cycles in the bi-variables analysis involved in the cross-wavelet mean-time spectrum with the WaveletComp version 1.0 statistical package in R. These results observed in the wavelet analysis showed the great complexity involved in the interaction of the different oceanic and atmospheric scales, with temporal scales defined as decadal and interdecadal, which present a synchronic variability of the solar activity in the Sunspots, being able influencing at the Ocean/Atmosphere system in the decadal temporal scales. The 11-year cycle in sunspots is well determined, but frequency oscillations of 2 to 5 years can also be referenced and accompany the solar activity variability. In the most recent period in the analyzed series, it presented periods with 2.66, 5.33, 10.66 and 21.33 years, for each observed period can be understood as ranges in which it is modulated by solar activity. The results found characterized a multidecadal variability with order of 10.66 and 21.33 years time scales.

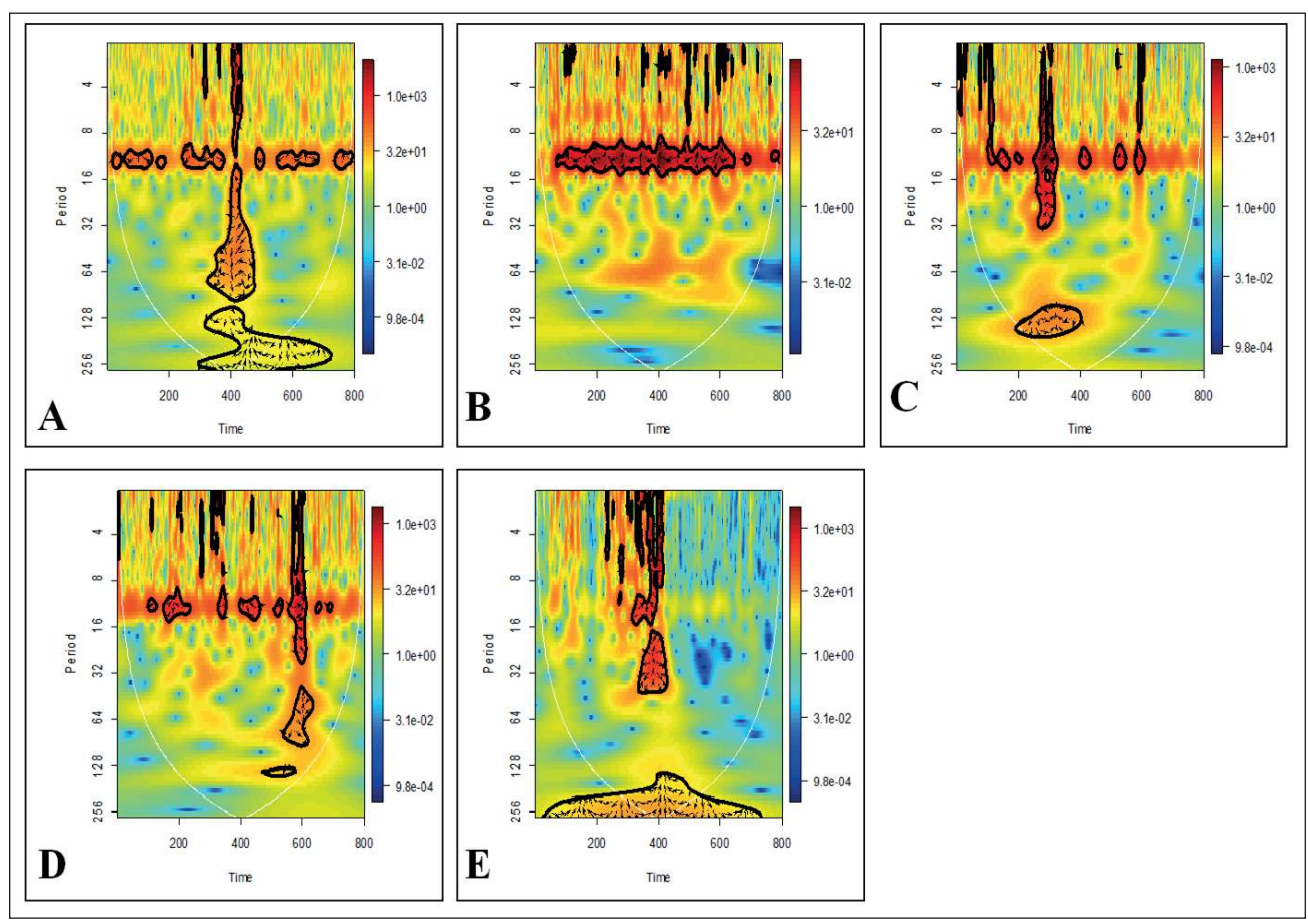

Figure 3 The time series periodic power wavelet spectrum of monthly total precipitation for the **A.** Belém; **B.** São Luiz; **C.** Fortaleza; **D.** Natal and **E.** Fernando de Noronha regions.

The total monthly precipitation spectrum was analyzed for locations from north to northeast Brazil, showed the same signals ranges but with differences due to continental location and the positioning and predominant meteorological systems influence of the atmosphere general circulation as example the Inter-tropical Convergence Zone (ITCZ) acts with spatial/temporal differences in the north and northeast Brazil regions. The ITCZ position is associated the Atlantic Meridional Mode (AMM) index is defined as the leading mode of non-ENSO coupled ocean/atmosphere variability in the Atlantic basin, in its negative or positive phase may characterize the northern or more southern location, influencing rainfall in the north-northeast region of Brazil. This situation will be studied in future work.

The existence of low frequency cycles in the time series allows us to analyze the occurrence of maximum and minimum events, since the different temporal scales involved may be constructively or not, and in future studies we will analyze overlapping events of multidecadal, intradecadal and seasonality in the time series, such as its influence on precipitation and other meteorological variables. Other works will be carried out in order to analyze the precipitation totals and their spatial variability, associating interaction structures oceanic/atmosphere general circulation.

# 5 Acknowledgements

The authores thank the support of the Instituto de Aeronáutica e Espaço (IAE).